\documentclass[showpacs,preprintnumbers,amsmath,amssymb]{revtex4}

\usepackage{dcolumn}
\usepackage{bm}

           \usepackage{graphicx}
           \usepackage{dcolumn}
           \usepackage{bm}
           \usepackage{epsfig}

\begin{document}

\title{Yang-Lee and Fisher zeros generalized on some far-from-equilibrium systems.}

\author{K. G.~Sargsyan\\[1mm]
{\small \textsl{Department of Theoretical Physics, Yerevan Physics
Institute,}}\\{\small Alikhanian Br. 2, 375036, Yerevan, Armenia
}\\[1mm]}
\begin{abstract}
A generalization of the Yang-Lee and Fisher zeros on
far-from-equilibrium systems coupled with two thermal baths is
proposed. The Yang-Lee zeros were obtained for minimal models which
exhibit complicated behavior in the context of the partition
function zeros and provide an analitycal treatment. This type of
models may be considered as a simplest one and analogous to Ising
model for equilibrium. The obtained distributions of generalized
Yang-Lee zeros show nontrivial behavior for these simple models.
\end{abstract}
\maketitle

More than five decades ago Yang and Lee proposed an approach to
clarify how singularities of the thermodynamic functions appears
within the canonical ensemble \cite{Yang}. They considered
liquid-gas transition and wrote down partition function in
grand-canonical description as a polynomial with respect to
fugacity. The main idea was an analysis of the fugacity in complex
plane. Complex roots of this polynomial with respect to fugacity
cross the real positive semi-axis in the singularity points in the
thermodynamic limit and single out the points of the phase
transition. Further, these roots in the complex fugacity plane are
called Yang-Lee zeros. After Yang and Lee pioneer work many papers
are devoted to this description of the phase transitions.

The approach of Yang-Lee zeros was used to describe spin system
where complex fugacity in grand-canonical ensemble was replaced by
the complex value of $e^{-\beta h}$ with $h$ as a magnetic field. It
was shown that the density of the Yang-Lee zeros is in a close
relation with the critical exponents describing the phase transition
and may be used as a measure of the strength of the transition.
Besides the Yang-Lee zeros so-called Fisher zeros were invented in
complex temperature plane \cite{Fisher}. Their properties are more
dependent on the particular system. For the historical details and
useful references on Yang-Lee and Fisher zeros see review
\cite{Bena}. Here we describe only some general aspects.

It was shown by Binek \emph{et al.}\cite{Binek} that the Yang-Lee
zeros are not fully and only a theoretical concept. They proposed an
experimental way to measure the Yang-Lee edge singularity exponents
from the isothermal magnetization data in 2D Ising ferromagnet.

In summary, the partition function zeros are rather well
investigated in case of the systems in equilibrium and provides us
a clue in understanding of the phase transitions. One can measure
some quantities which are directly related to the partition
function zeros and get some experimental evidence. The question is
how generalize an approach of Yang-Lee zeros to nonequilibrium
systems and when such a possibility exists.

A standart approach to describe the behavior of the nonequilibrium
system is to derive master equation and determine the rates of the
processes in it. After a long time, the system may settle in a
nonequilibrium steady-state. The exact meaning of the "long time"
depends on the rates of processes in it. According to initial
state and values of the parameters describing the system there may
be different steady states and the transition from one of them
into one another may be considered as an analogy to equilibrium
phase transition. The one possibility to generalize the partition
function approach is a consideration of the nonequilibrium models
which allow the transfer matrix description. As it is proposed by
Arndt \cite{Arndt} some nonequilibrium models which are useful for
the description of the driven diffusive systems, traffic flow,
biological transport and other processes give us a possibility to
invent the partition function and its zeros. In these models we
can express the stationary probability distribution as a trace
over some algebra elements, formulate the concept of partition
zeros and show the direct relation between steady-state transition
and the distribution of the partition function zeros. However, it
is possible to derive the distribution of the Yang-Lee zeros
without having the exact expression for the partition function.
One can consider the zeros of a steady-state normalization factor
in the complex plane of the transition rates \cite{Blythe}.

In this work one another possibility to generalize the the partition
function zeros is proposed. Let us consider thermodynamics of two
temperature systems with different time-scales and Hamiltonian
$H(\sigma,s)$ depending on fast variables $\sigma$ coupled with
thermal bath at temperature $T_{\sigma}$ and slow variables $s$
coupled with bath at temperature $T_{s}$. The system may be
considered as a minimal model having easily controlled
nonequilibrium properties \cite{minim}. The one of the real examples
may be NMR/ESR physics, where one of the baths is realized by weak
dipole interactions and the second one is a lattice temperature
\cite{gold}.

In such a system one can expect that after some long time system
settles in a nonequlibrium steady-state with heat currents between
the baths. This state is called steady adiabatic \cite{Allahv}.
Although the stationary distribution is far from Gibbsian, one can
derive the Gibbs like corresponding stationary distribution. As it
is shown in \cite{All} this type of nonequilibrium systems has
complicated and unusual behavior from equilibrium point of view.
There may be nonequilibrium phase transitions with no equilibrium
counterpart. Also, it is possible for latent heat to be negative in
presence of conflicting interactions \cite{All}. We expect that the
properly generalized partition function zeros may violate some usual
equilibrium statements too.

According to \cite{All} and taking into account a huge difference
of rates between fast and slow variables one can write down the
conditional probability
\begin{equation}
P(\sigma|s)=\frac{1}{Z(s)}e^{-\beta
H(\sigma,s)},Z(s)=Tr_{\sigma}e^{-\beta H(\sigma,s)}.\label{prob}
\end{equation}
Here we assume that on the times relevant to $\sigma$, $s$ does
not change and the conditional probability Eq. (\ref{prob}) is
Gibbsian. The next step is to derive $P(s)$. This may be done by
averaging the force $\partial_{s}H(\sigma,s)$ acting on $s$ over
the distribution $P(\sigma|s)$
\begin{equation}
P(s)=\frac{1}{Z}Z^{\frac{T_{\sigma}}{T_{s}}}(s),Z=Tr_{s}e^{-\frac{1}{T_{s}}}F(s),\label{prob1}
\end{equation}
where $F(s)=-T_{\sigma}lnZ(s)$ is the conditional free energy and
the common probability is $P(\sigma,s)=P(s)P(\sigma|s)$.

It is possible to obtain the steady state distribution by
minimizing the free energy
\begin{equation}
F=-T_{s}lnZ.\label{free1}
\end{equation}
Eq. (\ref{free1}) shows the way to define the partition function
zeros. It may be done like the equilibrium one, after replacing the
equilibrium partition function by $Z$ in common equilibrium
description. As it is in equilibrium, one may suppose that there is
a possibility of the far-from-equilibrium phase transition between
steady states in thermodynamic limit \cite{All}, when the
probability distribution is quasi-Gibbsian (\ref{prob1}) (as it is
for Gibbsian). The example of phase transition is demonstrated in
\cite{All}. However, this simple example is still mathematically
rather complicated. The motivation of usefulness of the partition
function zeros and their application for the indication of the phase
transition points in case of the quasi-Gibbsian distribution is the
same as it is in equilibrium. The main purpose of this work is not
to solve complicated models demonstrating phase transitions and
rather difficult for exact solution. Our aim  is to investigate the
properties of the partition function zeros for the models providing
analytical solution and compare the results with equilibrium
analogs.

Let us discuss the simplest models which may be treated
analytically and define the partition function zeros on those
examples. As such simple systems one can take the one-dimensional
Ising model with nearest-neighbor interactions and couplings or
magnetic fields as fast variables
\begin{equation}
H=-\sum_{i=1}^{N}J_{i}s_{i}s_{i+1}-\sum_{i=1}^{N}
h_{i}s_{i}.\label{ham}
\end{equation}
For the simplicity, let the couplings $J_{i}$ are the fast
variables and take values $\pm1$. In that case one may write down
for $Z(s)$ the following expression
\begin{equation}
Z(s)=Tr_{\{J_{i}\}}e^{-\frac{1}{T_{J}}H}=2^{N}\prod_{i}\cosh(\frac{s_{i}s_{i+1}}{T_{J}})e^{\frac{h}{T_{J}}s_{i}},\label{z1}
\end{equation}
where $T_{J}$ is a temperature of the thermal bath coupled to fast
variables. From Eq. (\ref{z1}) we get for $Z$
\begin{equation}
Z=Tr_{\{s_{i}\}}Z(s)^{\frac{T_{J}}{T_{s}}}=TrV^{N},\label{z2}
\end{equation}
where the transfer-matrix
\begin{equation}
V=[2\cosh(\frac{s_{i}s_{i+1}}{T_{J}})]^{\frac{T_{J}}{T_{s}}}e^{\frac{h}{T_{s}}\frac{(s_{i}+s_{i+1})}{2}},\label{transn}
\end{equation}
is introduced. Here we assume the periodic boundary conditions and
that the system consists of $N$ spins. The transfer-matrix $V$ is
symmetric and one can write $Z$ with respect to eigenvalues of
$V$, as it is done in equilibrium case for one-dimensional models
\begin{equation}
Z=\lambda_{1}^{N}+\lambda_{2}^{N}.\label{z3}
\end{equation}
The condition $Z=0$ leads to
\begin{equation}
\lambda_{2}=\lambda_{1}e^{\imath
\varphi},\varphi=\frac{(2k+1)\pi}{N},k=0,...N-1.\label{lambda}
\end{equation}
One of the examples of this procedure and its more detailed
description in equilibrium case may be found in \cite{Glumac}.

As a generalization of the equilibrium Yang-Lee zeros one can
consider solution of Eq. (\ref{lambda}) with respect to
$\mu=e^{\frac{h}{T_{s}}}$. However, the model with fast random
couplings is too trivial. After taking into account symmetry of
the function $\cosh$ and some simple algebra one gets the set of
the Yang-Lee zeros as $\mu=\pm \imath$. A similar set of values
$\mu =(\pm \imath)^{T_{h}/T_{s}}$ would be obtained if we
generalize Yang-Lee zeros as $\mu=e^{\frac{h}{T_{J}}}$.

In case of fast magnetic fields $h_{i}=\pm h$ the corresponding
transfer-matrix is
\begin{equation}
V=\left[2\cosh\left(\frac{h(s_{i}+s_{i+1})}{2T_{h}}\right)\right]^{\frac{T_{h}}{T_{s}}}e^{\frac{Js_{i}s_{i+1}}{T_{s}}}\label{transn1}
\end{equation}
It is easy to obtain for the eigenvalues of Eq. (\ref{transn1})
the following expression
\begin{equation}
\lambda_{1,2}=\left[2\cosh\left(\frac{h}{T_{h}}\right)\right]^{\frac{T_{h}}{T_{s}}}e^{\frac{J}{T_{s}}}\pm
2^{\frac{T_{h}}{T_{s}}}e^{-\frac{J}{T_{s}}} .\label{eigen}
\end{equation}
Here the natural and most simplest choice of the generalized
Yang-Lee zeros is solution of Eq. (\ref{lambda}) with respect to
$e^{\frac{h}{T_{h}}}$. For both $e^{\frac{h}{T_{h}}}$ and
$e^{\frac{h}{T_{s}}}$ the general assumption is a complex magnetic
field and the corresponding temperature is chosen according to
details of model. It is easy to obtain the second type of the
Yang-Lee zeros if the first one is known and vice versa.

\begin{figure}
          \begin{center}
          \includegraphics[width=7.5cm]{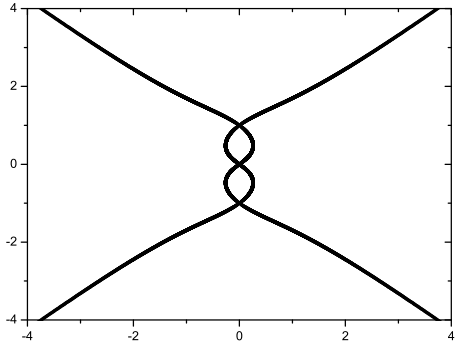}
          \caption{\label{1} The distribution of generalized Yang-Lee zeros in
          complex $\mu=e^{\frac{h}{T_{h}}}$ plane, where $J=0.5$ and $T_{h}=10, T_{s}=5$.}
          \end{center}
        \end{figure}

Denoting by $\mu=e^{\frac{h}{T_{h}}}$ and
$a=2^{\frac{T_{h}}{T_{s}}}e^{-\frac{J}{T_{s}}}$ we obtain
\begin{equation}
\left[\left(\mu+\mu^{-1}\right)\right]^{\frac{T_{h}}{T_{s}}}e^{\frac{J}{T_{s}}}+
a=
(\left[\left(\mu+\mu^{-1}\right)\right]^{\frac{T_{h}}{T_{s}}}e^{\frac{J}{T_{s}}}-a)e^{\imath\varphi}.\label{equat}
\end{equation}
Using condition on $\varphi$ (\ref{lambda}) one can derive a
finite set of independent equations corresponding to all possible
values of $\varphi$. These equations for generalized Yang-Lee
zeros are transcendental in general. One can consider a simple
choice of rational ratio $\frac{T_{h}}{T_{s}}$ or for more
simplicity we assume it to be a natural number. It is possible for
the natural ratio $\frac{T_{h}}{T_{s}}$ to solve the set of
equations numerically for even a large numbers of
$N\approx10^{5}$. The obtained distribution may be considered as
to be close to the thermodynamic limit. However, the fast magnetic
fields model has an obvious analytical solution in thermodynamic
limit. Near the thermodynamic limit one can treat $\varphi$ as a
continuous parameter, running over the interval $(0,2\pi)$ due to
relation $N\rightarrow\infty$. One can derive from Eq.
(\ref{equat}) the following relation
\begin{equation}
\left[\left(\mu+\mu^{-1}\right)\right]^{\frac{T_{h}}{T_{s}}}=2^{\frac{T_{h}}{T_{s}}}e^{-2\frac{J}{T_{s}}}\frac{e^{\imath\varphi}+1}{e^{\imath\varphi}-1}
.\label{rel}
\end{equation}
As an example, if the ratio is $\frac{T_{h}}{T_{s}}=2$ one gets
two simple quadratic equations and the corresponding solutions are
\begin{equation}
\mu_{1,2,3,4}=\frac{\pm c \pm \sqrt{c^{2}-4}}{2},\label{sol}
\end{equation}
where
$c=2e^{-\frac{J}{T_{s}}}\sqrt{\frac{e^{\imath\varphi}+1}{e^{\imath\varphi}-1}}$.
The corresponding distribution is represented in Fig. \ref{1}. Four
solutions of Eq. (\ref{sol}) form a complicated picture. This
behavior is rather different from equilibrium one, when for a simple
Ising-type model one gets partition function zeros lying on a unit
circle. Also, it is easy to detect from Eq. (\ref{sol}) that for
$\varphi\rightarrow 0$ the both four solutions are
$\mu_{1,2,3,4}\rightarrow\infty$.  In equilibrium case the usual
condition on Yang-Lee edge singularity points is $\varphi=0$ (see
for examples \cite{Glumac,Ruben} ).
\begin{figure}
          \begin{center}
          \includegraphics[width=7.5cm]{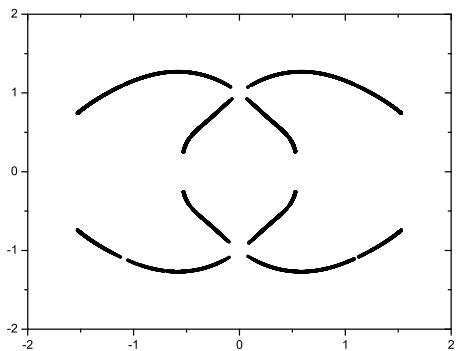}
          \caption{\label{2} The distribution of generalized Yang-Lee zeros in
          complex $\mu=e^{\frac{\gamma}{T_{\gamma}}}$ plane, where $N=10^{5}$ and $T_{\gamma}/ T_{h}=2$.}
          \end{center}
        \end{figure}
The second step one can do is to consider model with fast
variables $\gamma_{i}=\pm \gamma$, slow variables $s_{i}=\pm 1$
and Hamiltonian as
\begin{equation}
H=-\sum_{i=1}^{N}\gamma_{i}s_{i}s_{i+1}-\sum_{i=1}^{N}
\gamma_{i}s_{i}.\label{hams}
\end{equation}
The corresponding transfer-matrix is
\begin{equation}
V=[2\cosh(\frac{\gamma
s_{i}s_{i+1}}{T_{\gamma}}+\frac{\gamma}{T_{\gamma}}\frac{(s_{i}+s_{i+1})}{2})]^{\frac{T_{\gamma}}{T_{s}}}.\label{transn2}
\end{equation}
It is straightforward to define the generalized Yang-Lee zeros for
this model, as partition function zeros with respect to
$\mu=e^{\gamma/T_{\gamma}}$. After writing down characteristic
equation for transfer-matrix and obtaining the eigenvalues it is
simple to get the Yang-Lee zeros due to Eq. (\ref{lambda}). After
some numerical calculation within Mathematica we obtained a highly
nontrivial distribution for such a simple model (see Fig. \ref{2}
). The behavior of the zeros (Fig. \ref{2}) also violates the
circle theorem.

In order to generalize the Fisher zeros one can introduce two
types of the Fisher zeros according to two temperatures. In
equilibrium limit, when the ratio $\frac{T_{h}}{T_{s}}$ is equal
to one those two types of the Fisher zeros coincide. Simple
algebra shows that the same pointless set we obtain for the Fisher
zeros in case of fast couplings. For the model with fast magnetic
fields one need to solve more complicated equations compared with
those for Yang-Lee zeros.

Summarizing, the basic approach to generalize the partition
function zeros for non-equilibrium systems coupled to two thermal
baths is proposed. The described zeros show nontrivial behavior
and need to more complete investigation. This approach of the
generalized partition function zeros may be used to analyze phase
transitions by numerical and analytical methods as it is done in
equilibrium case. As author supposes, it is interesting to discuss
the properties of nonequilibrium Yang-Lee and Fisher zeros in more
details and for more complicated systems.

 Author is grateful to A. E. Allahverdyan for bringing attention to
this field and N. Ananikian for useful discussions. This work was
partly supported by grant ANSEF PS-condmatth-521.

\end{document}